\documentclass[final,9pt]{iiisconf}
\usepackage{multicol}
\usepackage{epsfig}
\usepackage{amsmath}

\begin{document}
\title{Topology and Geometry of Online Social Networks}
\author{%
\bf Dmitry Zinoviev\\%
\bf Mathematics and Computer Science Department, Suffolk University\\%
\bf Boston, 02114, USA%
}
\date{}
\maketitle

\section*{ABSTRACT}
In this paper, we study certain geometric and topological properties of online social networks using the concept of density and geometric vector spaces. ``Moi Krug'' (``My Circle''),  a Russian social network that promotes the principle of the ``six degrees of separation'' and is positioning itself as a vehicle for professionals and recruiters seeking each others' services, is used as a test vehicle.
\vskip\baselineskip\noindent
{\bf Keywords:} Online Social Network, Friend, Density, Metric Space, Vector Space, Chebyshev Space.

\section{OVERVIEW}
Social networks made their way into the Internet and became an important (if not the most important) part of the socially oriented Web in the beginning of the 21$^\mathrm{st}$ century  (Friendster 2002~\cite{boyd2004}, MySpace 2003, LinkedIn 2004, FaceBook 2004, Yahoo 360$^\circ$ 2005)~\cite{novak06,weaver2008}. 

Initially, the predominant language of the social Web was English or the languages based on the Latin-1 character set, especially Spanish and German. The participation of Russian-speaking Web users in these social networks was limited due to the relative underdevelopment of the Russian segment of the Internet and the fact that most Russian teenagers and young adults, who constitute the core of social networks, felt uncomfortable, if not unpatriotic, to communicate in the English language. One noticeable exception was the bilingual Russian diaspora, mainly in the USA, Western Europe, and Israel, which successfully got integrated into the major English-speaking social networks and is not the topic of this study.

The rapid proliferation of accessible high-speed Internet access in Russia in the early 2000s~\cite{alizar05} removed the first obstacle, while the emergence of the new generation of Russian Web programmers and reasonably cheap Web hosting made it possible to develop Russian-language social networks, which offered Russian as the default and the only interface language. Thus, the new social networks formed an isolated cluster limited to Russian-speaking participants, which historically tend to be ethnic Russians or the nationals of the C.I.S. and the Baltic states.

The most sizable Russian online social networks to date are ``Odnoklassniki'' (``Classmates,'' http://odnoklassniki.ru), ``V Kontakte'' (``In contact,'', http://vkontakte.ru), and ``Moi Krug'' (``My Circle,'' http://moikrug.ru), further referred to as MKOSN. ``Odnoklassniki'' aims at classmates, schoolmates, former coworkers, and ``brothers-in-arms,'' mostly helping people to reestablish lost contacts and later switch to an alternative mode of communications, such as e-mail or phone. It is organized by schools, universities, army units, major companies, and---interestingly---popular vacation sites. In this sense, it is not a place where people socialize, but a ``lost-and-found'' directory. Until recently, ``Odnoklassniki'' allowed users to post at most one photograph to their profiles and did not have the concept of ``friends\footnote{The concept of ``Odnoklassniki'' changed substantially in the Winter 2007--2008: despite poor interaction support, the site now has all essential features of a social network.}.''

``V Kontakte'' is a recent copycat of FaceBook. It has virtually the same functionality, except for the lack of an open interface.

On the contrary, ``friendship'' is the core concept of the MKOSN. The MKOSN introductory page emphasizes the principle of the ``six degrees of separation,'' or ``six handshakes,'' as it is known in Russia. The network is organized not by professional or academic affiliations, but by individuals and their ``circles.'' Each MKOSN member is surrounded by his or her ``first circle'' of the immediate friends, the ``second circle'' of the friends of the friends, and the ``third circle'' of the friends of the friends of the friends (these three circles are implemented in the MKOSN explicitly).

Despite the organizational structure based on the notion of proximity rather than on occupation, MKOSN positions itself as a vehicle for professionals and recruiters seeking each others' services. The recruiters and HR specialists with hundreds of ``friends'' in their inner circles have influenced the network's statistics. However, the magnitude of this influence cannot be easily estimated.

MKOSN is a relatively ``young'' network. It has been put to service in November 2005. To the best of our knowledge, as of July 2007 the network has approximately 166 thousand members (see \appendixname~\ref{acquire-sec}). This makes it a unique testing ground for various network exploration algorithms: because of the network's small size, it is possible to apply slow algorithms (such as $O(N^2)$ and even NP-complete) to the entire network, rather than to its subset, thus avoiding the burden of proving that the selected subset adequately represents the network.

This paper studies the internal structure and the peculiarities of ``Moi Krug,'' both from the mathematical and psychological points of view.

\section{MACROSCOPIC PARAMETERS\label{macro}}

A typical online social network consists of a giant core $\Gamma$---a subnetwork that contains the majority of connected members---and smaller marginal components not connected to the core and to each other~\cite{white2001}. At the moment of acquisition, the MKOSN was a tiny social network: it had only 166 thousand nodes in the giant core. This number is consistent with the estimate made in~\cite{senkut07} several months earlier (0.1 mln. users).

Various macroscopic parameters of online social networks have been analyzed, e.g., in~\cite{jamali06} and~\cite{novak06}.

\subsection*{Node Degree}

\begin{figure}[tb!]\centering
\epsfig{file=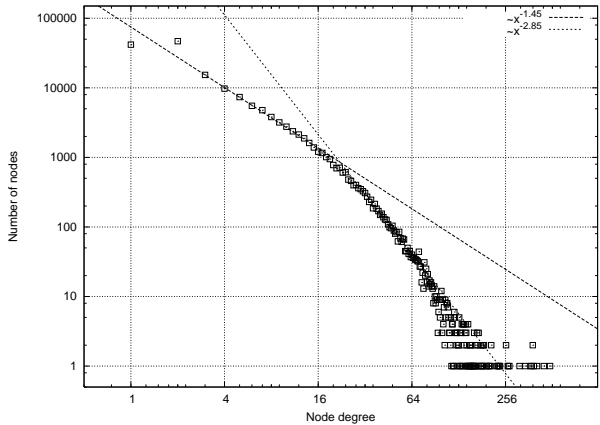,height=\columnwidth,angle=-90}
\caption{\label{pareto}Node degree distribution in MKOSN is a double Pareto distribution. The $x$-axis is the degree and the $y$-axis is the number of nodes at this rank.}
\end{figure}

One of the most fundamental properties of a non-directed graph is the node degree distribution. It has been observed that the node degrees of major social networks are distributed according to the double Pareto law. The MKOSN node distribution is not an exception (Figure~\ref{pareto}).

Reed~\cite{reed01} suggests that the break point in the double Pareto distribution is due to the fact that the age of the observed nodes is distributed exponentially: the ``young'' nodes are on the left, and the ``old'' nodes are on the right. (If the nodes' ages were distributed uniformly, all nodes would be ``young,'' and the distribution would follow the power law.) The fact that the MKOSN degree distribution has a break point may be an indication of the presence of at least two generations of nodes and members: ``senior'' members (those with 25 or more ``friends;'' these members have been in existence, say, since the establishment of the social network), and ``junior''  members. According to Figure~\ref{pareto}, the ``senior'' members constitute $\approx$3.6\% of the giant core size.

If seniority is indeed a reason for having the break point and the ``senior'' nodes were added to the network at about the same time (in other words, they were the core of the original social network), then they would be strongly connected in a sense that there would be at least one other ``senior'' node in the immediate vicinity of any other ``senior'' node. The MKOSN analysis confirms the hypothesis: out of 5,900 ``senior'' nodes in the giant core, only 66 (1\%) do not have any ``senior'' neighbors. Moreover, the subgraph of the ``senior'' nodes is dense: most ``senior'' nodes have 8--10 ``senior'' neighbors, and the average number of ``senior'' members in a neighborhood of a ``senior'' node is 19.

Unfortunately, the absence of the dynamic data from MKOSN does not allow us to verify the hypothesis about the generational origin of the double Pareto distribution.

\subsection*{Path Length}
\begin{figure}[t!]\centering
\epsfig{file=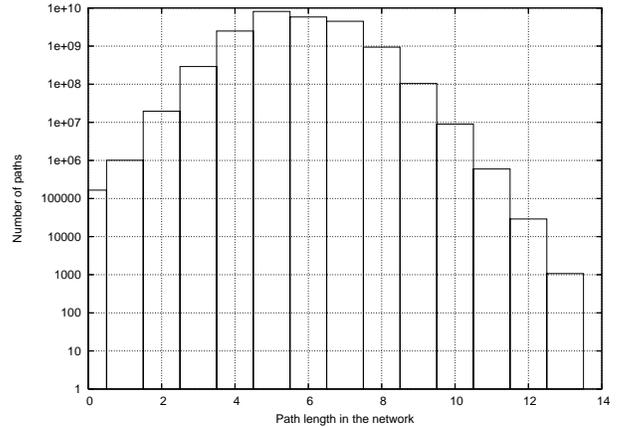,height=\columnwidth,angle=-90}
\caption{\label{path}Path length in the network}
\end{figure}

The second important macroscopic parameter of a graph is the distribution of node-to-node path lengths. The MKOSN claims that it is built around the principle of the ``six degrees of separation.'' As it turns out (Figure~\ref{path}), there are on the order of $10^{10}$ paths in the network. The longest path (the diameter of the network) is 14 hops long, which is more than twice the length of the legendary ``six degrees.''

On the other hand, the mean path length is six hops: the majority of the nodes are indeed ``six degrees'' apart from one another. 

\section{FINE STRUCTURE OF THE MKOSN}
The macroscopic parameters presented in Section~\ref{macro} help us little to understand the internal structure of the social network. We propose three finer-grain approaches (two topological and one geometric) to the social network analysis.

\subsection*{Macroscopic topology\label{macrotopology}}
One can see from Figure~\ref{pareto} that 25\% of the MKOSN nodes have the degree of 1. They are part of the giant core, but are loosely connected to the rest of the network. The corresponding network members have been apparently introduced to the MKOSN by their more active friends, but had neither time nor desire to expand their contact lists.

\begin{figure}[t!]\centering
\epsfig{file=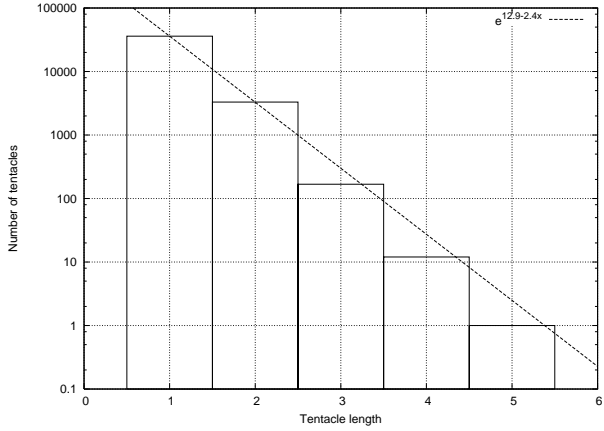,height=\columnwidth,angle=-90}
\caption{\label{tentacles}Path length in the tentacles}
\end{figure}

A more careful analysis reveals that $\approx$2\% of the network nodes have the degree of two and are connected to at least one other node that has the degree of one or two. Such nodes form ``tentacles'' that expand from the denser part of the network ``outward.''\footnote{The word ``outward'' is quoted because so far there is no ``in'' or ``out'' direction in the network.} Thus, a tentacle is a chain of network members, each of which is a ``friend'' of the next one and the previous one, except for the last marginal member (the loner), who has only one ``friend.'' The tentacles' lengths are distributed exponentially with the mean length of 1 hop (Figure~\ref{tentacles}). The exponential nature of the distribution suggests that the probability of a loner to add another ``friend'' is a constant. 

The cyberpsychological nature of the tentacles is currently unclear.

The nodes that do not belong to the tentacles form the dense core $\Delta$ ($\Delta\subset \Gamma$). The size of the MKOSN dense core is 123 thousand nodes.

\begin{figure}[b!]\centering
\epsfig{file=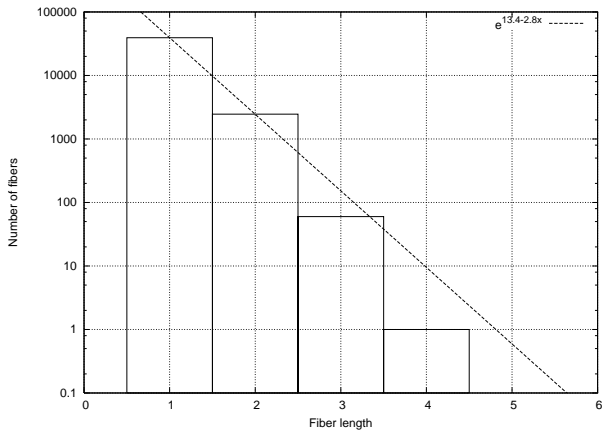,height=\columnwidth,angle=-90}
\caption{\label{fibers}Fiber length in the dense core}
\end{figure}

It is tempting to check if the dense core is uniformly dense or it has ``cavities,'' possibly crossed by thin ``fibers.'' A fiber is very similar to a tentacle, except that the ``loner'' end of a fiber is connected back to the dense core (using the topological terminology, the dense core with a fiber is a sphere with a handle). The MKOSN has 42,000 fibers with an average length of  2 hops (Figure~\ref{fibers} shows the distribution of the number of inner nodes in a fiber, which is one less the number of hops). The lengths of the fibers are distributed exponentially, too.

The fibers are probably formed when the loners add ``friends'' with the same constant probability as during the tentacle construction, but one of the newly added ``friends'' is already a member of the dense core. The cyberpsychological nature of the fibers is probably the same as that of the tentacles.

\begin{figure}[t!]\centering
\epsfig{file=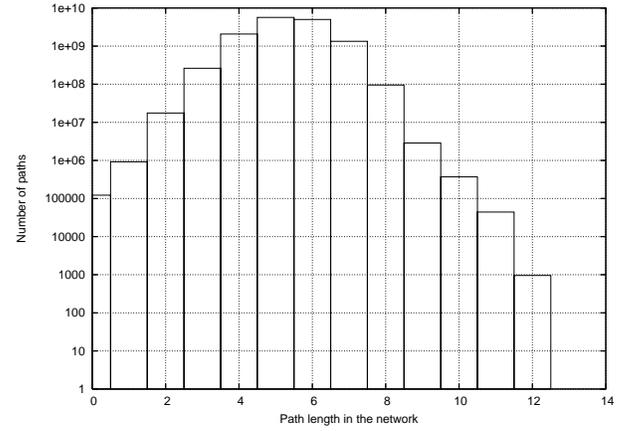,height=\columnwidth,angle=-90}
\caption{\label{path-core}Path length in the dense core}
\end{figure}

\begin{figure}[b!]\centering
\epsfig{file=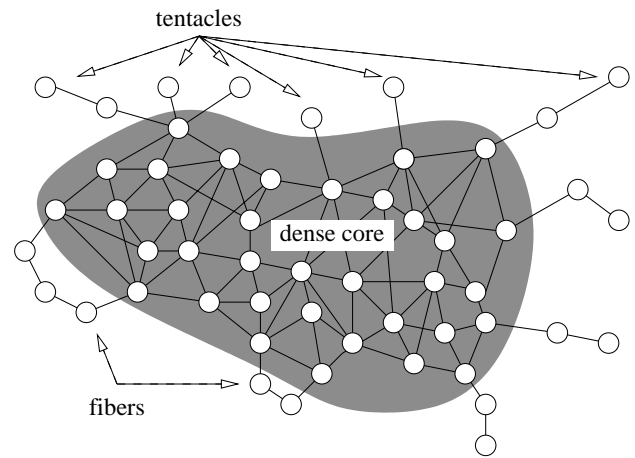,width=\columnwidth}
\caption{\label{structure}The macroscopic structure of the MKOSN}
\end{figure}

As the result of the macroscopic topological study, we identified the fine structure of the MKOSN (Figure~\ref{structure}): the dense core that amounts to 74\% of the giant core, the tentacles, and the fibers. Since the tentacles are very simple in nature, they can be eliminated from further consideration. The distribution of the node-to-node path lengths in the dense core (including the fibers, but excluding the tentacles) is shown in Figure~\ref{path-core}. The mean path length in the dense core is five hops---one hop less than in the giant core, meaning that the majority of ``regular'' network members are actually even closer to each other than the network claims.

\subsection*{Mesoscopic Topology}
The macroscopic topological study does not address the structure of
the dense core. In particular, we do not know if the core is uniformly
dense and if it has inner and outer (boundary) nodes. The answer can
be given by exploring the mesoscopic (medium-range) network topology.

The idea of describing the mesoscopic topology of a social network
numerically is not new. An overview of major proposed
mechanisms---cliques, n-clans, and k-plexes---is given
in~\cite{jamali06}. Our approach considers the social network as a
continuous medium (which is somewhat acceptable if the number of
network nodes is large) and is based on the density study.


With no vector space associated with the network, we have to redefine the density so that the new definition does not depend on any vector properties. In particular, the new definition cannot use the concept of volume. The ``classical'' definition $\rho\left(x_0,\ldots x_D\right)=\frac{\partial N\left(x_0,\ldots x_D\right)}{\partial V}$, where $N$ is the number of nodes,  is ruled out.

We can compare a social network to a crowd of people. The crown is dense around an individual if the individual has many neighbors; otherwise, the crowd is sparse (technically speaking, it is not a crowd). Apparently, the number of network neighbors (the node degree) of a node $\aleph$ can serve as a reasonable estimate of the social network's density in the vicinity of $\aleph$: $\rho\left(\aleph\right)=\mathrm{degree}\left(\aleph\right)$.

If the proposed definition of density is used, then the MKOSN is not uniform, and the density follows the double Pareto distribution that has already been discussed. However, the graph in Figure~\ref{pareto} does not reflect the spacial density distribution.


\begin{figure}[t!]\centering
\epsfig{file=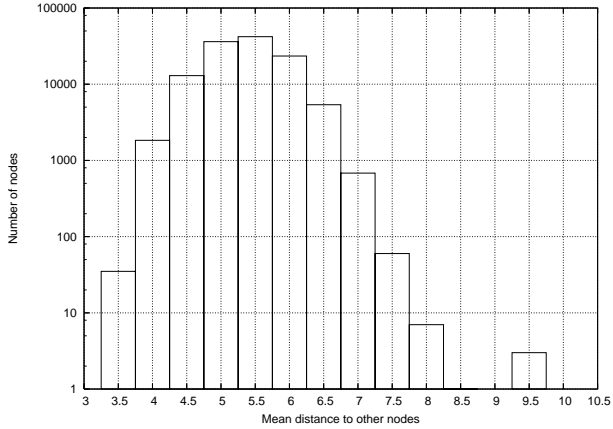,height=\columnwidth,angle=-90}
\caption{\label{radius-a-core}Depth distribution in the dense core (larger depths correspond to  the nodes that are closer to the ``boundary'')}
\end{figure}


\begin{figure}[b!]\centering
\epsfig{file=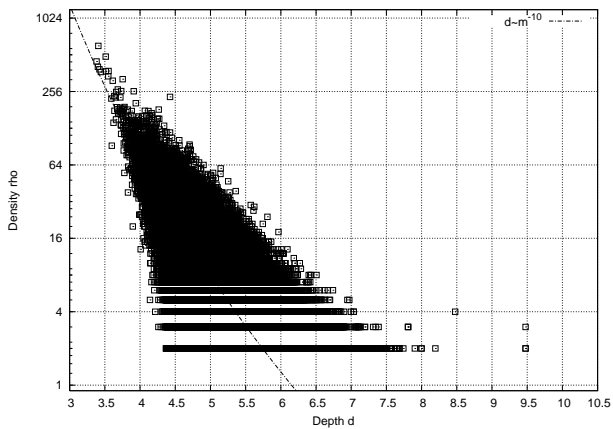,height=\columnwidth,angle=-90}
\caption{\label{radius-core}Dense core density $\rho$ as a function of depth $d$}
\end{figure}

To identify the ``inner'' and the ``boundary'' nodes of $\Delta$, we will use the observation that in an spherically symmetric $D$-dimensional object, the mean distance $\bar{m}$ from any point P to all other points of the object reaches the minimum if P is at the center. Any displacement of P from the center increases $\bar{m}$, and the maximum is reached on the boundary of the object. In particular, if our object were a uniform sphere, then for the point that is at the distance $r$ from the center, $\bar{m}=a\left(D\right)\sqrt{b^2\left(D\right)+r^2}$.

We define the distance between two nodes P and Q in the dense core as the minimum length of the paths connecting P and Q. The depth of P, $d\left(P\right)\equiv\bar{m}\left(P\right)$, is the mean distance from P to all other nodes in $\Delta$. The distribution of the depths in $\Delta$ is shown in Figure~\ref{radius-a-core}. The mean depth $\bar{d}=5.5$ equals the mean path length in the dense core (Figure~\ref{path-core}). It is reasonable to assume that ``boundary'' nodes lie on the right of the chart, and the ``inner'' nodes are in the middle and on the left. The figure shows that if our hypothesis about the spherical symmetry is correct, then the majority of the nodes are concentrated in the middle layer, with very few nodes on the true ``boundary'' and in the ``center.'' Speaking in other words, the overwhelming majority of the dense core members are ``average people,'' with very few marginal and socially popular members.

Now we are ready to combine the ``depth cues'' and the density information and plot $\rho$ against $d$ (Figure~\ref{radius-core}). It follows from the graph that the dense core is dense in the center and sparse at the outskirts: for the well-connected members (high-$\rho$), it is easier to reach the rest of the dense core (low-$d$).

By redefining the local density $\rho(\aleph)$, we can explore one more interesting aspect of the MKOSN. It is known that people are either socially active (socially popular, extroverts) or socially passive (marginals, introverts). A socially popular person has a lot of ``friends,'' which are not necessarily popular by themselves---at least not as popular as the person himself or herself. We can identify socially popular members and marginals by comparing the density at $\aleph$ and in its neighborhood.

\begin{figure}[t!]\centering
\epsfig{file=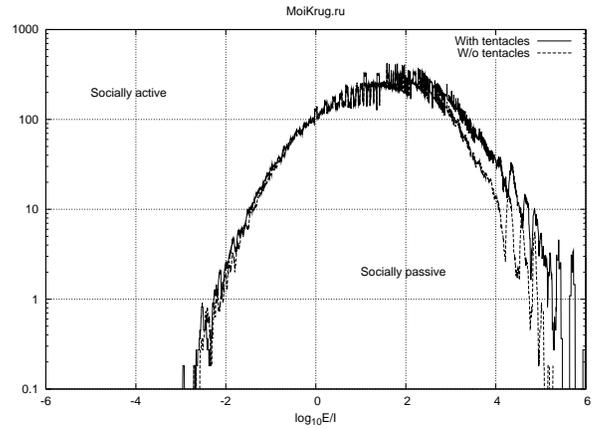,height=\columnwidth,angle=-90}
\caption{\label{ie}Quantitative personality $\Pi=\log_{10}E/I$ of the MKOSN members}
\end{figure}

Let $\epsilon\left(\aleph\right)$ be the set of all neighbors of $\aleph$ in $\Delta$---the first circle of $\aleph$. Then
\begin{equation} 
\rho_\epsilon\left(\aleph\right)\equiv\frac{\sum\limits_{a\in\epsilon\left(\aleph\right)}\rho\left(a\right)}{|\epsilon\left(\aleph\right)\!|}
\end{equation}
is the average density of the first circle. Let $E/I\equiv\rho_\epsilon\left(\aleph\right)/\rho\left(\aleph\right)$. Then $\Pi\equiv\log_{10}E/I$ is the quantitative personality---the measure of the social activity of a member. Positive $\Pi$ means that the ``friends'' of the member on average have more ``friends'' than the member, i.e., that member is socially passive, or a marginal. Negative $\Pi$ identifies members that are socially popular and have more ``friends'' than the members of their first circles.

The distribution of $\Pi$ for the dense core of the MKOSN is shown in Figure~\ref{ie}. The distribution is skewed to the right: the marginals outnumber the popular members in the dense core by the factor of 9. The ratio is even larger for the entire MKOSN (10:1).

Finally, we can draw some conclusions about the likelihood of a high-$\Pi$ (marginal) member to be in touch with a low-$\Pi$ (popular) member and the other way around. Table~\ref{lei} shows the fraction of marginal, neutral, and popular ``friends'' in the first circle of a MKOSN member.

\begin{table}[t!]\centering
\begin{tabular}{l|ccc}\hline
$\aleph$&\multicolumn{3}{c}{First circle of $\aleph$}\\\cline{2-4}
&Popular  & Neutral &Marginal\\\hline
Popular  & 26\% & 2\% & 72\%\\
Neutral   & 24\% & 4\% & 72\%\\
Marginal & 43\% & 3\% & 53\%\\\hline
\end{tabular}
\vskip0.5\baselineskip
\caption{\label{lei}Personalities in the MKOSN}
\end{table}

The table suggests that the popular and neutral network members tend to cluster with the more marginal (or less popular) members, while the marginals do not have clear preferences.

\section{GEOMETRY}

The topological studies of online social networks focus on defining their structure, as well as the boundaries and the inner areas. However, they do not consider the positions of the nodes (network members). 

It is tempting to construct a multidimensional vector space (perhaps even a linear space) that has the social network nodes as points. The coordinates of the nodes in such space may be related to the social properties of the underlying network or to the psychological properties of the network members. In this section, we attempt to elaborate the geometry of the MKOSN.

The graph of a social network induces a discrete metric space M, where the distance between two points (nodes), P and Q, is the minimum length of the paths connecting P and Q. The metric function $d_m\left(P,Q\right)$ is implicitly defined. We want to embed this
space into a D-dimensional vector space with minimal distortion. In other words, we want to assign D-tuples of coordinates to each point P in M: $P\rightarrow X_P=\left(x_P^0,x_P^1,\ldots,x_P^{D-1}\right)$---and a metric function $d_v\left(X,Y\right)$ so that $\forall P,Q\in M: P=Q\rightarrow |d_m\left(P,Q\right)/d_v\left(X_P,Y_Q\right)-1|<\epsilon$. The value of $\epsilon$ is called the distortion of the embedding~\cite{linial2002} and ideally should be infinitely small. 

As a first approximation, we propose to use the following vectorization procedure: Let us enumerate all nodes, and let $M_i$ be the node number i in M. Then for a node $P\in M$, let $x_P^i\equiv d_m(P,M_i)$. In other words, the i'th coordinate of P is the distance from P to the i'th node (we call the i'th node a reference node, or reference point). In particular, $x_P^i=0\iff P=M_i$. 

The metric function is based on the Chebyshev distance:
\begin{equation}
d_v(X,Y)\equiv\max_i\left(|x_P^i-y_Q^i|\right).
\end{equation}
It is not hard to see that $\forall P,Q: d_m(P,Q)=d_v(X_P,Y_Q)$. Thus, the newly constructed space is a non-distorting embedding of M.

Unfortunately, the dimensionality of the new space is too high: even for the relatively small MKOSN, there are 166,000 dimensions, which is probably well beyond any practical use. We will use the modification of the Quine-McCluskey method~\cite{quine1952} to reduce the dimensionality.

In any general network, some of the dimensions are dependent. For example, in a standalone tentacle it's enough to define one linear coordinate, no matter how long the tentacle is. Several dimensions are dependent if removing all of them but one does not change the metric function on M.

Let's form  matrix $Z[D\left(D-1\right)/2\times D]$ such that for $i>j$:
\begin{equation}
z^{ij}_k=
\begin{cases}
0&\text{if $|x^i_k-x^j_k|<\max(0,d(P_i,P_j)-T)$},\\
1&\text{if $|x^i_k-x^j_k|\ge\max(0,d(P_i,P_j)-T)$},
\end{cases}\label{zmatrix}
\end{equation}
where tolerance $T\ge 0$. T=0 gives the exact solution. 

A column $z_k$ in Z corresponds to the point $P_k$. We say that $z_k$ covers row $z^{ij}$ if $z^{ij}_k=1$. If column $z_k$ covers row $z^{ij}$, then it is an essential implicant: removing point $P_k$ from the set of reference points distorts the distance between $P_i$ and $P_j$ by more than T hops.

If a point is an essential implicant, we add it to the set E of essential implicants and remove all rows from Z the are covered by $z_k$. The procedure is repeated until the matrix Z has no more rows. The points that are not in the set E are not the reference points of the new vector space.

In our experiments, we could easily reduce the number of dimensions from 5 to 2 with no distortion and from 5 to 1 with the distortion of 25\% (T=1). Unfortunately, the storage complexity of the proposed algorithm ($O\left(D^3\right)$) makes it hard to use even for modestly sized social networks.

\section{CONCLUSION}
In the paper we investigated several topological and geometric approaches to the structural studies of online social networks in general and of the ``Moi Krug'' OSN in particular. We introduced the concepts of the dense core and the local density and analyzed the density distribution within the dense core. We further used the density to identify socially popular and socially marginal network members. We attempted to embed the topological metric space induced by the social network graph into a vector space and concluded that the embedding is either impractical due to the huge dimensionality of the resulting space or is computationally expensive and still not very practical. We conclude that the exploratory mechanisms based on topological properties are promising and preferable over the geometric mechanisms.

\setlength{\textheight}{218mm}
\appendix
\section{APPENDIX: NETWORK ACQUISITION\label{acquire-sec}}
\begin{figure}[b!]\centering
\epsfig{file=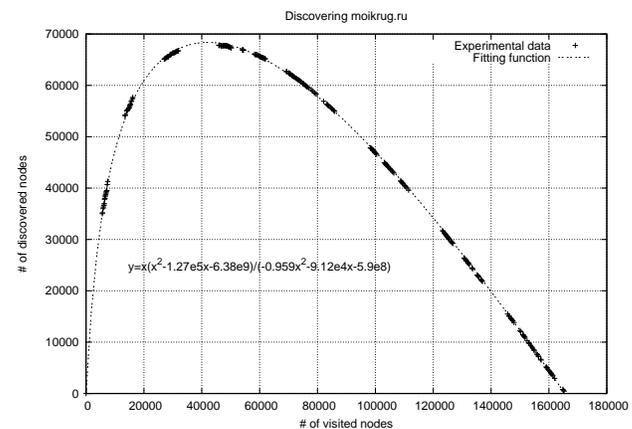,height=\columnwidth,angle=-90}
\caption{\label{acquire}Number of discovered (but not processed) nodes $D$ vs the number of processed nodes $P$: experimental data and an empirical graph}
\end{figure}

Due to the fact that the administration of MKOSN does not disclose the total number of the participants, and the ``rumor-based'' estimates variate from 80,000 to 400,000 members, it was important to develop a mechanism that would allow us to learn the approximate network size in advance. In case of a large number, we would have to limit our research to a subset of the network, rather than to the entire network.

The total size of the network $S$ is the sum of the number $P$ of nodes that have been already discovered and processed, the number $D$ of nodes that have been discovered, but not processed, and the unknown number $X$ of nodes that have not been seen yet:
\begin{equation}
S=P+D+X\label{sdpx}
\end{equation}

The processed nodes can be compared to the interior of the explored subspace $C_e$ of the space $C$, while the unprocessed nodes can be compared to its boundary. Processing $\Delta P$ nodes leads to the discovery of $\Delta D$ new nodes by following the links to these nodes (on average, $L$ ``external'' links per node), and to simultaneously moving $\Delta P$ nodes from the set of discovered nodes into the set of processed nodes: $\Delta D=+L\Delta P-\Delta P$. Therefore, $L=(\Delta D/\Delta P+1)\approx D'+1$. Intuitively, when $P\gg X$, then $X\approx DL\approx D\left(D'+1\right)$, and (\ref{sdpx}) becomes:
\begin{figure}[t!]\centering
\epsfig{file=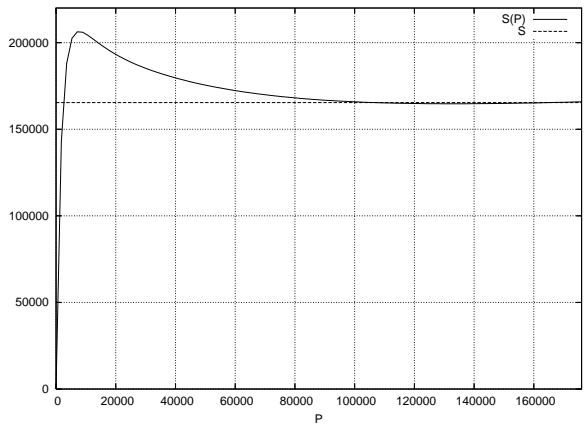,height=\columnwidth,angle=-90}
\caption{\label{s}Estimated and actual network size at various stages of the network acquisition}
\end{figure}
\begin{equation}
S\approx P+D+\left(D'+1\right)D\label{sp-nodiffur}
\end{equation}

Assuming that ideally $S$ should not depend on $P$, i.e., $S'=0$, Eq.~(\ref{sp-nodiffur}) can be rewritten as a second-order differential equation:

\begin{equation}
DD''+\left(D'+1\right)^2=0\label{sp-diffur}
\end{equation}

This equation does not have a closed-form solution. In our case, the acquisition curve can be very closely approximated by the following fractional-rational function (Figure~\ref{acquire}):

\begin{equation}
D=a_0P\frac{P^2+a_1P+a_2}{P^2+a_3P+a_4}
\end{equation}

This function also closely matches a corresponding numerical solution of (\ref{sp-diffur}).

By combining the experimental values of $P$ and $D$ and the evaluated value of $D'$, we can estimate the total network size at the early stages of network acquisition (Figure~\ref{s}). The difference between the predicted and actual sizes is less than 10\% for $P>35,000$, which means that the network size can be estimated fairly well after the acquisition of only 20\% of the nodes.

\section*{ACKNOWLEDGMENT}
The author is thankful to Dr. Boris Mirman, Prof. Pra\-deep Shukla (Mathematics department, Suffolk University), and Prof. Sukanya Jay (Psychology department, Suffolk University) for their enlightening discussions that resulted in my deeper understanding of the geometry and topology of social networks and the cyberpsychology of the network members. Special thanks to Ms. Vy Duong for proofreading and editing the manuscript.

\bibliographystyle{acm}
\bibliography{cs}

\end{document}